\documentclass[prb,twocolumn,showpacs,amsmath,amssymb,preprintnumbers,superscriptaddress]{revtex4}
\usepackage{dcolumn}
\usepackage{bm}
\usepackage{graphicx}
\usepackage{subfigure}
\usepackage{color}

\begin{document}

\title{Magnetic behavior, Griffiths phase and magneto-transport study in 3$d$ based nano-crystalline double perovskite Pr$_2$CoMnO$_6$}

\author{Ilyas Noor Bhatti}\email{inoorbhatti@gmail.com}\affiliation{Department of Physics, Jamia Millia Islamia University, New Delhi - 110025, India.}
\author{Rabindra Nath Mahato}\affiliation{School of Physical Sciences, Jawaharlal Nehru University, New Delhi - 110067, India.}
\author{Imtiaz Noor Bhatti}\email{inbhatti07@gmail.com}\affiliation{School of Physical Sciences, Jawaharlal Nehru University, New Delhi - 110067, India.}
\author{M. A. H. Ahsan}\affiliation{Department of Physics, Jamia Millia Islamia University, New Delhi - 110025, India.}

\begin{abstract}
 Double perovskite (DP) oxide material receive extensive research interest due to exciting physical properties with potential technological application. 3$d$ based DP oxides are promising for exciting physics like magnetodielectric, ferroelectric, Griffith phase etc., specially Co/Mn DPs are gaining much research interest.  In this paper we present the study of magnetic phase and transport properties  in nano-crystalline  Pr$_2$CoMnO$_6$ a 3$d$ based double perovskite compound. This material shows a paramagnetic (PM) to ferromagnetic (FM) phase transition below 173 K marked by a rapid increase in magnetic moment due to spin ordering. We found divergence in inverse magnetic susceptibility ($\chi$$^{-1}$) from Curie weiss behavior around 206 K which indicates the evolution of Griffiths phase before actual PM-FM transition. We found that the Griffiths phase suppressed with increasing applied magnetic filed. For the understanding of charge transport in this material we have measured temperature dependent electrical resistivity. Pr$_2$CoMnO$_6$ is a strong insulator where resistivity increase abruptly below magnetic phase transition. To understand the effect of magnetic field on transport behavior we have also measured the magnetoresistance (MR) at different temperatures. Sample shows the negative MR with maximum value $\sim$22 $\%$ under applied magnetic field of 50 kOe at 125 K. MR follows quadratic field dependency above $T_c$ however below $T_c$ the MR shows deviation from this field dependency at low field. 
\end{abstract}

\pacs{75.60.Jk, 75.75.-c, 75.47.Lx}

\maketitle
\section{Introduction}
Double perovskites with the general formula A$_2$BB$^{\prime}$O$_6$ (where the B and B$^{\prime}$ sites are occupied by transition metal ions and the A site is occupied by rare earth ions) have been revitalized due to its varied magnetic and electrical properties and its importance for potential application in information technology and different spintronics applications like spin-based sensors, multiple-state memory elements, relaxor ferroelectricity, magnetodielectric capacitors etc.\cite{ado, rov, son, choudhury, kawa, nair, masud, ali, ullah, rto} In particular, 3$d$ based double perovskite compounds form fascinating class of  material, and have receive great attention of researchers due to their unusual physical properties. For instance, high temperature ferromagnetic La$_2$NiMnO$_6$ is a potential candidate for Piltier cooler.\cite{ado} Further, lattice driven ferroelectricity has recently been reported in La$_2$NiMnO$_6$ thin films\cite{shi}. Recently, Matte \textit{et al},  investigated the thin films of La$_2$NiMnO$_6$ and mentioned it as material for implication of magnetocaloric devices.\cite{matte} In ferromagnetic Y$_2$CoMnO$_6$ and Er$_2$CoMnO$_6$ pyroelectric effect is observed and attributed thermally stimulated depolarization currents.\cite{sco1, sco2} Spin glass state at low temperature and giant magnetocloric effect is observed in Gd$_2$CoMnO$_6$ which undergoes a ferromagnetic transition around 131 K. \cite{hoo, wang} Pr$_2$CoMnO$_6$ is a ferromagnetic insulator with Griffiths singularity above $T_c$ in PM state found in bulk samples.\cite{liu, lshi} The combination of transition metal ions at B- and B$^{\prime}$-site is interesting, their intersection via oxyen anion can lead to interesting physics in this double perovsktes. However, A-site rare earth cation greately influence the physical properties in these compounds.\cite{son}

Pr$_2$CoMnO$_6$ is a ferromagnetic material with $T_c$ $\sim$173 K.\cite{lshi, khemchand} It also shows strong insulating properties with very large magnetoresistance due to which it is believed that strong magnetodielctric effect arise. Spin phonon coupling and exchange bise effect has been reported in polycrystalline bulk Pr$_2$CoMnO$_6$.\cite{lshi} Interestingly, two magnetic transition has been seen in magnetic data at T$_{c1}$ $\sim$173 K due to ordered phase i.e. Co$^{2+}$-O-Mn$^{4+}$ results in super exchange interaction and long range FM ordering, where at much lower temperature T$_{c2}$ $\sim$142 K there occurs another transition due to disorder phase Co$^{3+}$-O-Mn$^{3+}$ which give rise to vibronic superexchange interaction in bulk Pr$_2$CoMnO$_6$. However, the disordered phase disappears with decreasing particle size as reported by Shi \textit{et al}.\cite{lshi} Its is interesting to note here that reported inverse magnetic susceptibility shows divergence around 210 K in bulk Pr$_2$CoMnO$_6$ which is a signature of Griffiths Phase.\cite{liu} It would be quite interesting to investigate GP in nano-crystalline Pr$_2$CoMnO$_6$ which has not been attempted yet. Formation of FM clusters above the long-range FM ordering temperature in PM state is characterized by zero spontaneous magnetization (M$_s$). This clustered state with local spin interaction in global paramagnetic matrix is often described as a Griffiths phase.\cite{griffiths, zho} Originally the model was proposed for randomly diluted Ising ferromagnets. However, now the appearance of GP is acceptable in many other manganite and double perovskite systems.\cite{mandal, jiang, ang, chan, zhou, mon, akp1, akp2} In such systems GP is characterized by looking into several features in magnetic data e.g.  the low field temperature inverse susceptibility $\chi$${^-1}$(T) would show deviation from Curie-Weiss (CW) behaviors marked by sharp downturn in inverse susceptibility data. This deviation would be suppressed in large magnetic fields due to the polarization of spins outside the clusters, magnetic susceptibility at low field would obey modified CW law  $\chi$$^{-1}$(T) $\propto$ (T- T$_{C}^{R}$)$^{1 - \lambda}$ in GP region where 0$<$$\lambda$$<$1 is susceptibility exponent and its value describe how strong is the Griffiths phase. Further,  from Arrott plot ($H/M$ vs $M^2$) $M_s$ can also be checked in this region to confirm the existence of GP.\cite{zho, akp1, akp2}

In this paper, we present synthesis and characterization of nano-crystalline Pr$_2$CoMnO$_6$ material. Synthesis of nano-crystalline sample was done by sol-gel method. The structural characterization is done by X-ray diffraction technique. The structural analysis shows that the sample is in single phase and crystallizes in orthorhombic structure with average crystallite size $\sim$17 nm. The magnetic properties are studied with the investigation of DC magnetization data. The sample shows a PM-FM phase transition around 173 K. The apperance of Griffith phase is observed above FM transition in paramagnetic state. The detail analysis of magnetization data above $T_c$ conforms the presence of Griffith singularity. We further studied the sample for charge transport, the temperature dependent resistivity shows the hopping of small polarons that gives the conduction mechanism. Magnetoresistance collected at different temperature shows a huge decrease in resistance in FM ordered state well below $T_c$. We found a maximum negative MR of 22 $\%$ at 125 K under 30 kOe field. The magneto resistance data shows a quadratic field dependency above $T_c$, whereas this behavior is limited to high field at low temperatures.  

\section{Experimental Detail}
Nano-crystalline sample of Pr$_2$CoMnO$_6$ was prepared by the sol-gel method. High purity ingredients Pr(NO$_3$)$_3$.6H$_2$O, Co(NO$_3$)$_2$.6H$_2$O and C$_4$H$_6$MnO$_4$.4H$_2$O in powder form were taken in stoichiometric ratio and were dissolved in water to form clear solution. Further 3:1 ratio of Citric acid is dissolved in water to obtain solution. Then two solutions were mixed in  a 500 ml beaker. To allow sufficient time for reaction the beaker with solution is filled with double distilled water up to 400 ml and then placed on magnetic stirrer with continuous stirring for 24 hour at 90 $^o$C to form gel. After evaporation of liquid the remains in the beaker are heated at 300 $^o$C. After 3 hours of heating the collected residue was grounded for one hour to form fine powder. Then the powder was  sintered  at 900 $^o$C for 12 hours. The phase purity of samples was checked by x-ray diffraction (XRD) using Rekigu made diffractometer. The XRD data was collected at room temperature in 2$\theta$ range 10$^o$-90$^o$ with step size of 0.02$^o$ and scan rate of 2$^o$/min. Crystal structural analysis was performed by rietveld refinement of XRD data using Fullprof program. Scanning electron microscopy image was obtained for particle size analysis and morphology. The DC magnetization data and electrical transport measurements was measured in the Physical Properties Measurement System by Cryogenic Inc.

\begin{figure}
	\centering
		\includegraphics[width=8cm]{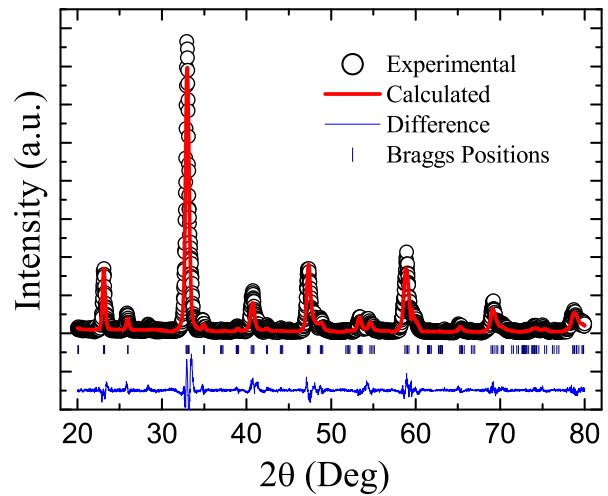}
	\caption{(color online) Powder X-ray diffraction pattern along with rietveld refinement is shown for Pr$_2$CoMnO$_6$ sample.}
	\label{fig:Fig1}
\end{figure}

\section{Result and Discussion}

\subsection{Structural Characterization}
For structural characterization of nano-crystalline Pr$_2$CoMnO$_6$ sample powder x-ray diffraction data was recorded. The detailed structural analysis  for Pr$_2$CoMnO$_6$ was done by rietveld refinement of XRD data using fullprof program.  Fig.1 shows the XRD patterns along with Rietveld refinement. In this samples, black open circles represents experimental data, red solid line is the calculated pattern and blue thin line is the difference, Braggs positions are represented by navy blue bars. The rietveld refinement of XRD pattern is fine with acceptable fitting perimeters goodness of fit $\chi^2$ and R$_{wp}$/R$_{exp}$ ratio as 2.25 and 1.38 respectively. These values are considered to be acceptable for good fitting. \cite{imtiaz1, imtiaz2} Rietveld refinement shows the samples crystallize in single phase and is chemically pure. Pr$_2$CoMnO$_6$ adopt orthorhombic crystal structure with \textit{Pbnm} space group.  The lattice parameters $a$ $b$ and $c$ obtained from structural analysis are 5.446(5) $\AA$, 7.695(1) $\AA$ and 5.403(4) $\AA$ respectively. We have used Debye Scherer's formula D = k$\lambda$/$\beta$cos$\theta$ to calculate the crystallite size. Here k is grain shape factor $\lambda$ is x-ray wave length, $\theta$ and $\beta$ are Bragg's reflections and full width at half maxima fo the corresponding peak respectivelly. The obtained average crystallite size for present compound is $\sim$17 nm.

\begin{figure}
	\centering
		\includegraphics[width=8cm]{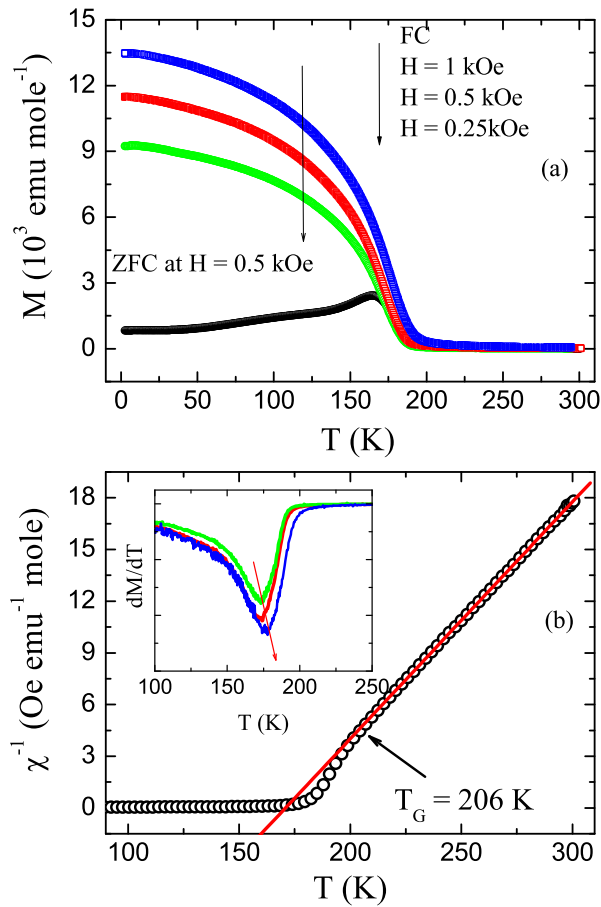}
	\caption{(color online) (a) DC Magnetization measured in ZFC and FC protocol, ZFC is measured at an applied field of  0.5 kOe and FC data is collected at different applied magnetic field and is shown as function of temperature for  Pr$_2$CoMnO$_6$. (b) Temperature dependent inverse susceptibility $\chi$$^{-1}$ $=$ $(M/H)$$^{-1}$ as deduced from magnetization data in Fig.2a. The solid line is due to Curie Weiss law Eq. 1. Inset shows dM/dT vs T plot.}
	\label{fig:Fig2}
\end{figure}

\begin{figure}
	\centering
		\includegraphics[width=8cm]{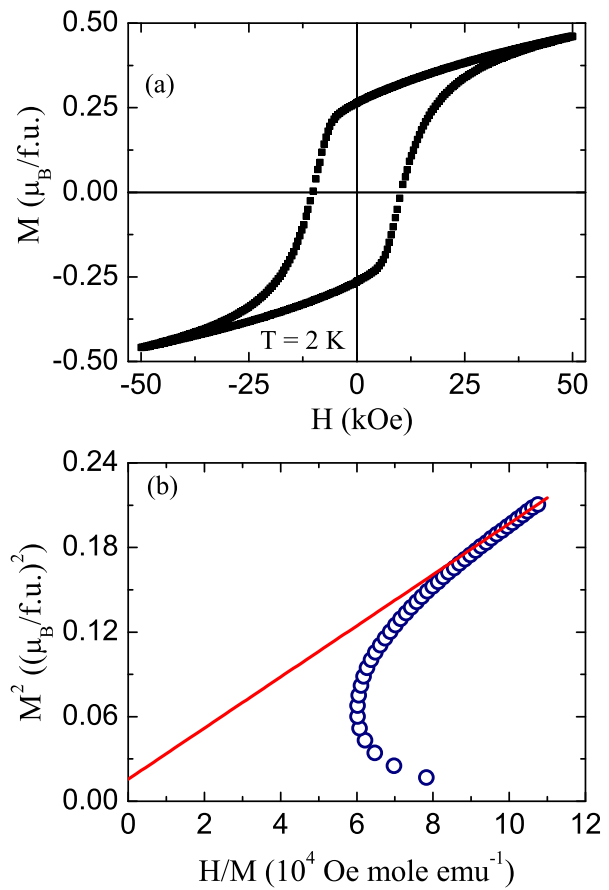}
	\caption{(color online) (a) Isothermal magnetization as function of applied field up to $\pm$50 kOe collected at 2 K is shown for Pr$_2$CoMnO$_6$. (b)  Arrott plot (M$^2$ vs H/M) as obtained from virgin curve of magnetization isotherm in Fig.3a,  solid line is high field extrapolation to obtain spontaneous magnetization (M$_s$).}
	\label{fig:Fig3}
\end{figure}

\textit{}\subsection{Magnetization Study}

Fig. 2a shows the temperature dependent magnetization data M(T) for Pr$_2$CoMnO$_6$ performed under zero-field cooled (ZFC) and field cooled (FC) protocol. $M_{ZFC}(T)$  data is collected at 0.5 kOe applied field, where as  $M_{FC}(T)$ is collected at three different applied magnetic fields viz. 0.25 kOe, 0.5 kOe and 1 kOe.  In Fig. 2a it is quite evident that on decreasing temperature the magnetic moment $M_{FC}$(T) shows sudden rise below 180 K which is an evidence of peramagnetic to ferromagnetic phase transition. We have identified the phase transition temperature by ploting $M_{FC}$(T) data as dM/dT vs T, where $T_c$ is marked by point of inflection as shown in the inset of Fig. 2b.  However, $M_{ZFC}$(T) data is rather interesting where with decreasing temperature it shows a hump near $T_c$ and with further decreasing temperature the moment decreases with two small kinks marked at T$_{M1}$ $\sim$110 K and T$_{M2}$ = $\sim$50 K as shown in Fig. 2a. It is evident from the Fig. 2a that there is a huge bifurcation between $M_{ZFC}$(T) and $M_{FC}$(T) data curves. Further for better understanding of the magnetic phase transition we have collected $M_{FC}$(T) data at two more fields, where we found that with increasing applied field, $T_c$ moves to higher temperature. The dM/dT vs T plot for all $M_{FC}$(T) data is plotted and is shown in inset Fig. 2b $T_c$ is marked by point of inflection in dM/dT vs T plot for $M_{FC}$(T) at 0.25 kOe, 0.5 kOe and 1 kOe applied field $T_c$ comes out to be $\sim$173 K, 174 K and 177 K respectively. 
 
\begin{table*}[th]
\caption{\label{label} Table shows the magnetic parameters transition temperature ($T_c$) Curie Wiess constant ($C$), effective magnetic moment ($\mu$$_{eff}$), Curie paramagnetic temperature ($\theta$$_{P}$), corecive force ($H_c$), remenant magnetization ($M_r$) and spontaneous magnetization ($M_S$) obtained from magnetic data in Fig. 2 and Fig. 3 are given for Pr$_2$CoMnO$_6$ sample.}
\begin{ruledtabular}
\begin{tabular}{ccccccc}
$T$$_c$ (K) &$C$ (emu K mol$^{-1}$ Oe$^{-1}$) &$\mu$$_{eff}$ ($\mu_B/$f.u.) &$\theta$$_{P}$ &$H$$_c$ (kOe) &$M$$_r$ ($\mu_B$/f.u.) &$M$$_S$ ($\mu_B$/f.u.)\\
\hline
173	&7.267  &7.60 &172.67 &11 &0.2633 &0.125 \\
\end{tabular}
\end{ruledtabular}
\end{table*}

To further understand the magnetic behavior in Pr$_2$CoMnO$_6$, we have plotted the temperature dependent inverse susceptibility ($\chi^{-1}$) for this material. The $\chi^{-1}(T)$ shows linear behavior at higher temperature, linearity in $\chi^{-1}$(T) is observed till 206 K. At this temperature there is a sharp downturn in $\chi^{-1}(T)$ which is a typical characteristic of Griffiths phase as discussed in next section. This downturn in inverse susceptibility at 206 K is marked as Griffiths temperature (T$_G$). We have fitted the inverse susceptibility ($\chi^{-1}$(T)) data above 210 K in paramagnetic state with curie Weiss Law shown in Fig. 2b, the solid  straight line is due to fitting with Eq. 1:

\begin{eqnarray}
 \chi = \frac{C}{T - \theta_P}
\end{eqnarray}

\begin{eqnarray}
 C = \frac{N_A \mu^2_{eff}}{3 k_B}
\end{eqnarray}

\begin{eqnarray}
 \mu_{eff} = \sqrt{\frac{3 N_A k_B}{\frac{d\chi^{-1}}{dT}}}
\end{eqnarray}

where $C$ is the Curie constant, $\mu_{eff}$ is the effective paramagnetic moment and $\theta_P$ is the Curie temperature. Fig. 2b shows that above 206 K, $\chi^{-1}(T)$ data can be well fitted with Eq. 1. Using the fitted parameter slope and intercept, we have calculated $C$, $\theta_P$ and $\mu_{eff}$ for Pr$_2$CoMnO$_6$. The experimental obtained value of $\mu_{eff}$ = 7.6 $\mu_B$/f.u. To understand such high value of effetive paramagnetic moment present in this compound we have attempted to calculate the theoretical value of $\mu_{eff}$. Since, in Pr$_2$CoMnO$_6$ besides Pr$^{3+}$ it has two other magnetic cations Co$^{2+}$ and Mn$^{4+}$  in ordered state where as due to some disorder present in the sample there may have other charge state of these cations viz. Co$^{3+}$/Mn$^{3+}$.  We have calculate the average effective moment $\mu_{eff}$ = $\sqrt{(^{Pr^{3+}}\mu_{eff})^2 + (^{Co^{2+}}\mu_{eff})^2 + (^{Mn^{4+}}\mu_{eff})^2}$. Where  $(^{Co^{2+}}\mu_{eff})$ and $(^{Mn^{4+}}\mu_{eff})$ were calculated considering high spin state for both Co$^{2+}$ (t$_{2g}^3$$\uparrow$ e$_g^2$$\uparrow$ $t_{2g}^1$$\downarrow$) and Mn$^{4+}$ (t$_{2g}^3$$\uparrow$ e$_g^0$$\uparrow$ $_{2g}^0$$\downarrow$) using $g\sqrt{S(S+1)}\mu_B$ spin-only systems, while for ($^{Pr^{3+}}\mu_{eff}$) = $g\sqrt{J(J+1)}\mu_B$ is used to calculate the $\mu_J$ for Pr$^{3+}$ {4f$^2$ $\uparrow$}cation since $J$ is consider as good quantum number for lanthanides and  Lande factor g = 0.8.  The average $\mu_{eff}$ = 7.06 $\mu_B$/f.u. which is less then the experimental observed value of $\mu_{eff}$ = 7.60 $\mu_B$ from Curie Weiss fitting of susceptibility data. The observed high value may be due to presence of disordered Co$^{3+}$/Mn$^{3+}$ cations with high spin state and large $\mu_{eff}$ Further, the experimentally obtained positive value of $\theta$$_P$ = 173 K  signifies strong ferromagnetic ordering present in the nano-crystalline Pr$_2$CoMnO$_6$. The magnetic parameter are calculated using fitting details given in Table-I.

Fig. 3a shows the isothermal magnetization $M(H)$ measured at 2 K in the range of $\pm$ 50 kOe for Pr$_2$CoMnO$_6$. The $M(H)$ curve exhibits a typical ferromagnetic behavior of this material. The $M(H)$ curve shows large hysteresis at 2 K with coercive field $H_c$ value $\sim$ 11500 Oe.  Further the sample shows large reminant magnetization (M$_r$) $\sim$0.235 $\mu_B/f.u.$.   In Fig. 3b, we have shown Arrott plot extracted from virgin curve of $M(H)$ data shown in Fig. 3a. Arrott plot ($M^2$ vs $H/M$) offers an effective tool to understand the nature of  magnetic state as the positive intercept on $M^2$ axis. Due to extrapolation of high field data of Arrott plot implies material exhibits spontaneous magnetization and hence FM state.\cite{akp1, akp2, imtiaz1, imtiaz2} From the intercept we calculated the value of spontaneous magnetization $M_S$ = 0.065 $\mu_B/f.u.$ which indicate a strong ferromagnetic state at low temperature. 
\begin{figure}
	\centering
		\includegraphics[width=8cm]{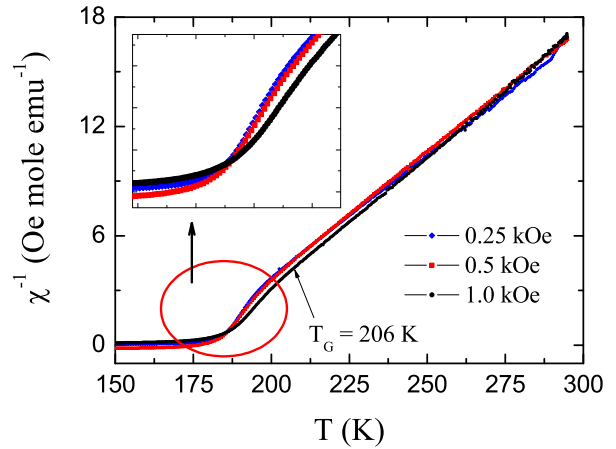}
	\caption{(color online) Inverse magnetic susceptibility as a function of temperature ($\chi$$^{-1}$ vs T) is shown for Pr$_2$CoMnO$_6$ collected at different fields. Inset shows the deviation of susceptibility from Curie Weiss behavior.}
	\label{fig:Fig4}
\end{figure}
\subsection{Griffiths Phase}
Fig. 2b shows the inverse magnetic susceptibility ($\chi^{-1}$) as a function of temperature. It is evident from the figure that with decrease in temperature from 300 K $\chi^{-1}$ obeys Curie Weiss law till 210 K, with further decrease in temperature a sharp downturn is observed in $\chi^{-1}$ vs $T$ plot which is a typical characteristic of Griffiths phase(GP).  At this point we need to confirm the existence of GP in this compound, for this we have measured dc magnetization data,  a different applied magnetic fields intensity $H$. Fig. 4a shows the $\chi^{-1}$ vs $T$ plot deduces from dc magnetization data as shown in Fig. 2a. The softening of downturn which is the signature of GP with increasing applied field from 0.25 kOe to 1.0 kOe conforms the formation of GP above $T_c$ in PM state of Pr$_2$CoMnO$_6$. Such feature, has also been reported in various other systems.\cite{akp1, akp2, magen} Griffiths phase is actually the appearance of finite-size ferromagnetic clusters in paramagnetic phases well above the $T_c$ where the spins are ferromagnetically correlated with in the clusters. However, the system as a whole would not develop any long range ordering and thus no spontaneous magnetization would appear. To further check the appearance of GP in this material we have measured $M(H)$ data at different temperatures. In Fig. 5 we have shown the Arrott plot ($M^2$ vs $H/M$) where intercept on the M$^2$ axis  by extrapolation of data from high field would gives, spontaneous moment ($M_s$). However,  Fig.5 shows representative Arrott plot at selective temperatures i.e. 158 K 164 K and 174 K below and above $T_c$. It is clear from Fig. 5 that Arrott plot for 158 K have positive intercept on $M^2$ axis and would give a finite $M_s$. However, Arrott plots above $T_c$ has negative intercept on $M^2$ axis and hence zero $M_s$. These experimental results confirms that in the GP regime the system does not have any $M_s$ related to long-range FM ordering. This further confirms that the downturn in in $\chi^{-1}$ in PM state is due to evolution of GP in  Pr$_2$CoMnO$_6$. Since, in GP the FM clusters appear with variable size with local FM ordering. In this senerio the magnetization is nonanalytic. It is usually found that in low field the magnetic susceptibility follow a power law behavior given by:\cite{griffiths, akp1, akp2}
\begin{figure}
	\centering
		\includegraphics[width=8cm]{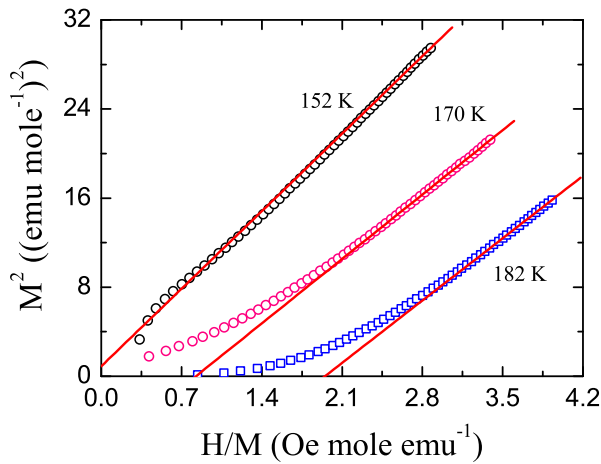}
	\caption{(color online) Magnetic isothermal virgin curve collected at different temperature is plotted as Arrott plott (H/M vs M$^2$) is shown for Pr$_2$CoMnO$_6$. The solid lines are extrapolation form high field.  }
	\label{fig:Fig5}
\end{figure}

\begin{figure}
	\centering
		\includegraphics[width=8cm]{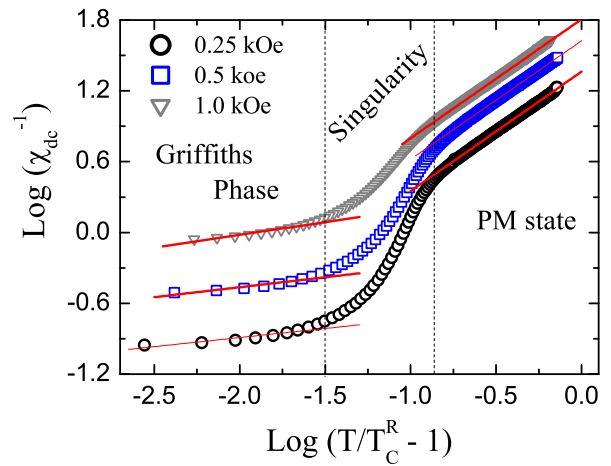}
	\caption{(color online) Log ($\chi$$^{-1}$) vs Log (T/T$_{C}^{R}$ is plotted following Eq. 4.}
	\label{fig:Fig6}
\end{figure}

\begin{eqnarray}
 \chi^{-1}(T) \propto (T- T_{C}^{R})^{1 - \lambda}
\end{eqnarray}

 where T$_{C}^{R}$ is the critical temperature of random FM where susceptibility tend to diverge and $\lambda$ (0$<$$\lambda$$<$1) is the magnetic susceptibility exponent. It is clear that the power law behavior in Eq. 4 is a modified Curie–Weiss law where the exponent $\lambda$ quantifies a deviation from Curie–Weiss behavior due to formation of magnetic clusters in the PM state above $T_c$. 

To further understand the Griffiths phase in nano-crystalline Pr$_2$CoMnO$_6$ we have plotted the magnetic susceptibility as Log ($\chi$$^{-1}$) vs Log (T/T$_{C}^{R}$ - 1) as shown in Fig. 6. We have fitted the data in Fig. 6 in both Griffith phase regime and paramagnetic state. It is worth noting here the determination of proper $\lambda$ is highly sensitive to the value of T$_C^R$ which means the incorrect value of $T_{C}^{R}$ in Eq. 4 can lead to unphysical fitting and erroneous determination of $\lambda$.  In order to estimate $T_{C}^{R}$ accurately we have followed the method reported earlier.\cite{akp1, jiang2} It is important to note here that the value of critical  temperature of random ferromagnetic $T_{C}^{R}$ always be greater than FM transition temperature $T_c$. In this method, first $T_{C}^{R}$ value is fixed to $T_c$ and $\lambda$$_{PM}$ is deduced from fitting of $\chi^{-1}$  at 0.25 kOe in PM state. Then the  $T_{C}^{R}$ replaced, thus so obtained $T_{C}^{R}$ is considered correct if in PM state the value of $\lambda$$_{PM}$ comes out to be close to zero.  This means that the $\chi^{-1}$ in PM state follows Curie Weiss behavior. The slope from straight line fitting in Fig. 6 for $\chi^{-1}$  at 0.25 kOe gives $\lambda$$_{GP}$ = 0.84 and $\lambda$$_{GP}$ = 0.023. The obtained values of $\lambda$$_{GP}$ for the nano-crystalline Pr$_2$CoMnO$_6$ is consistent with GP model. The larger value of $\lambda$ indicates that the Griffiths singularity is very strong for this material. Further, it is important to note that with increasing applied field we observed the positive value of $\lambda_{GP}$ viz. $\lambda_{GP}$ ($H$ = 0.25 kOe) = 0.84, $\lambda_{GP}$ ($H$ = 0.5 kOe) = 0.83 and $\lambda_{GP}$ ($H$ = 1 kOe) = 0.78. Even at higher field we observed that the value of susceptibility exponent is far from zero, thus keeping in mind this high value of $\lambda_{GP}$ we can say Griffiths-like phase remains at high field in nano-crystalline  Pr$_2$CoMnO$_6$. However, the suppression of the downturn in magnetic susceptibility is due to rising PM background and masking of FM signal.\cite{magen}  

\subsection{Electrical Transport and Magnetoresistance}

It is quite interesting to study charge transport in Pr$_2$CoMnO$_6$ nano-crystalline sample. This material is 3$d$ based transition metal oxide which is an insulator due to presence of strong electronic correlation effect.
Fig.7a shows the temperature dependent resistivity $\rho$(T) data for Pr$_2$CoMnO$_6$ sample collected at zero and 30 kOe applied magnetic field. It is obvious from the figure that the resistivity shows insulating behavior with very high resistivity. The room temperature resistivity $\rho$$_{300}$ is around then 100 $\Omega$-cm. The resistivity increases rapidly with decreasing temperature. We found that after 170 K there is drastic increase in resistivity by almost five order of magnitude and which exceed beyond measurable range below 130 K as shown in Fig.7a. Further, temperature dependent resistivity is measured in presence of 30 kOe magnetic field. The general feature of temperature dependent magnetoresistance data is similar to the normal $\rho$(T)  at zero field. Ovearall the sample remains in insulating phase. However, we observe decrease in resistivity at all temperatures i.e. negative magnetoresistance. Further, it is quite intriguing that the sample shows very large decrease in resistance below 170 K which is marked by PM-FM phase transition with $T_c$ $\sim$173 K. This shows that the ferromagnetic ordering of magnetic spins promote the flow of charge and hence resistivity decreases.

\begin{figure}
	\centering
		\includegraphics[width=8cm]{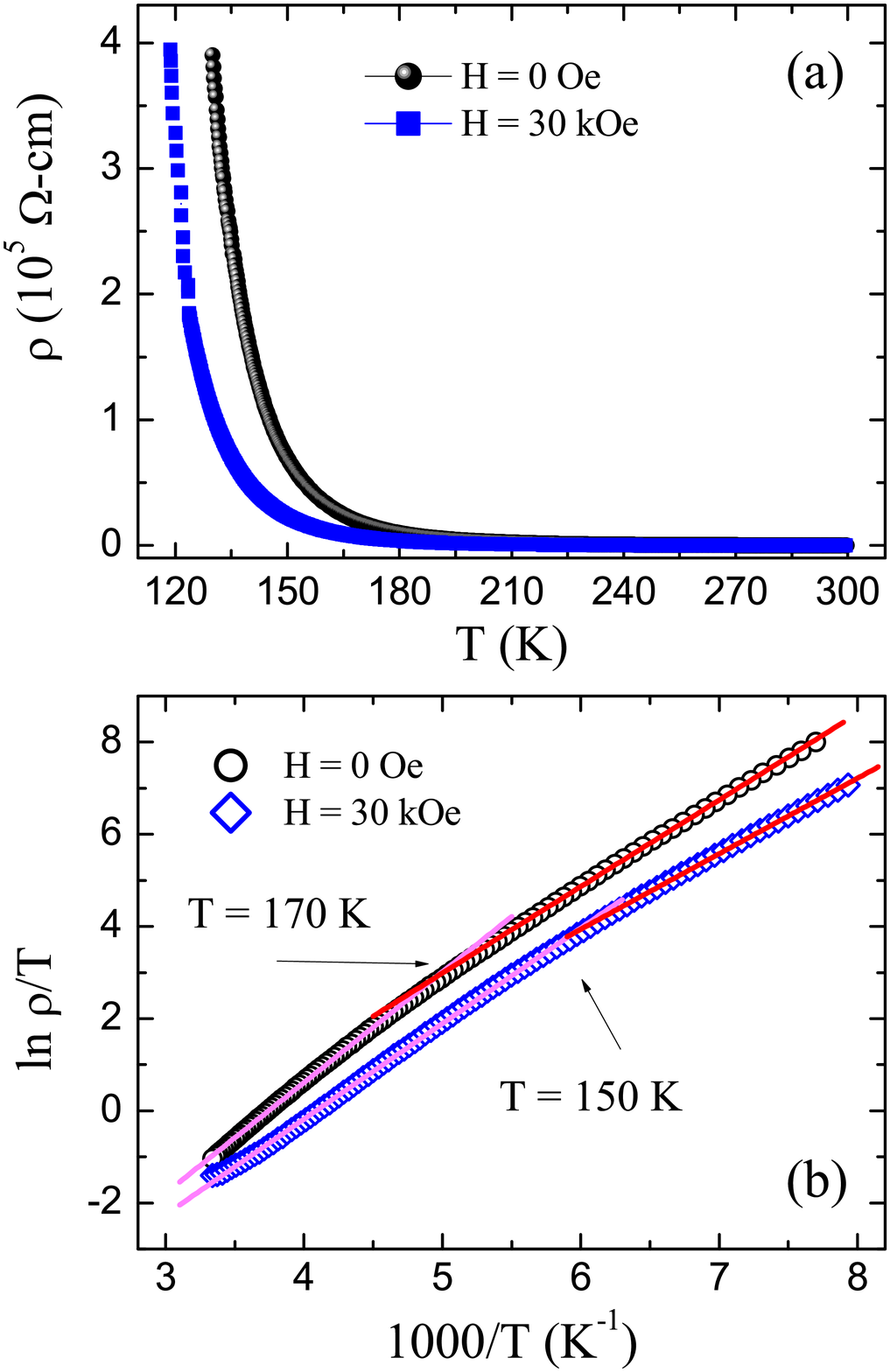}
	\caption{(color online) Temperature dependent resistivity $\rho$(T) data collected in 0 kOe and 30 kOe applied magnetic field is shown for Pr$_2$CoMnO$_6$ sample. (b) Temperature dependency of resistivity ln ($\rho/T$) vs T$^{-1}$ plot is shown. The solid lines are the least-square fitting due to Eq. 2.}
	\label{fig:Fig7}
\end{figure}
 
\begin{figure}[th]
	\centering
		\includegraphics[width=8cm]{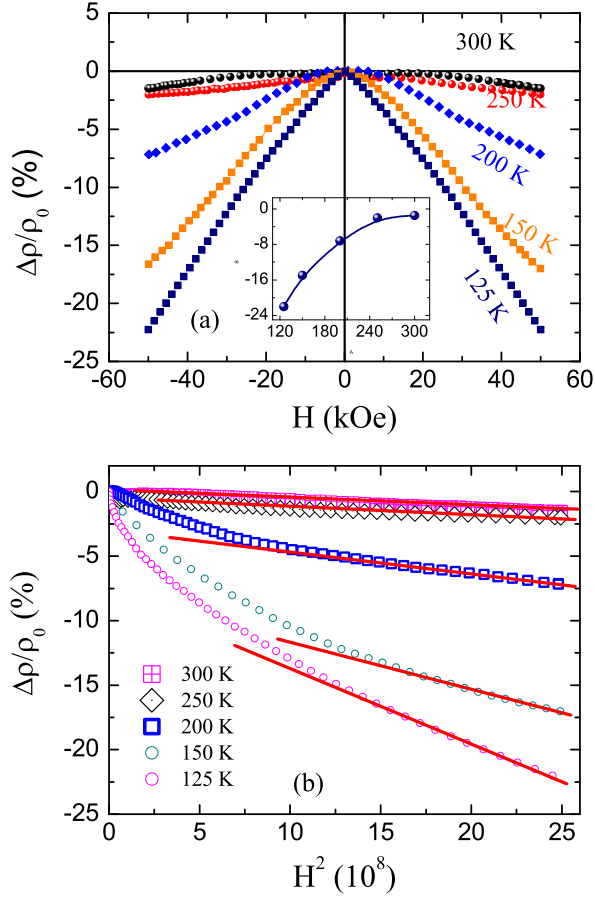}
	\caption{(color online) (a) Magnetoresistance collected at different temperature upto $\pm$50 kOe is shown for Pr$_2$CoMnO$_6$. (b) Quadratic field dependency of magnetoresistance (H$^2$ vs $\delta$$\rho$/$\rho$$_o$) is presented for Pr$_2$CoMnO$_6$ at different temperatures.The solid lines are due to Eq.3 }
	\label{fig:Fig8}
\end{figure}

To understand the  mechanism of resistivity we attempt to analyze resistivity data with different models. However we found that the resistivity follows inverse temperature dependency defined by following model:\cite{neifel}

\begin{eqnarray}
\rho \sim T exp{\left(\frac{\epsilon_\alpha}{k_B T}\right)}
\end{eqnarray}

 where $\epsilon$$_\alpha$ is activation energy and k$_B$ is the Boltzmann constant.

We analyzed the resistivity data for Pr$_2$CoMnO$_6$ following the above mentioned model. Fig.7b shows Log ($\rho$/$T$) vs $T^{-1}$ where the straight lines are due to Eq. 5. We observe that the resistivity data is well fitted in two ranges of temperature with slope change around PM to FM phase transition temperature ($T_c$).  The temperature dependent resistivity data collected at 0 kOe shows a slope change around 170 K where are data collected in applied magnetic field of 30 kOe undergo a slope change around 150 K as marked in Fig. 7b.

 To further understand the mechanism charge transport we have studied magnetoresistance (MR) in this material, the MR is the change of resistivity in the presence of applied magnetic field, described as:

\begin{eqnarray}
\frac{\Delta \rho}{\rho(0)} = \frac{\rho(H) - \rho(0)}{\rho(0)}
\end{eqnarray}
where $\rho(0)$ is the resistivity at zero field and $\rho(H)$ is the resistivity in presence of field.

Fig. 8a shows the isothermal MR measured at different temperatures. Sample shows negative MR i.e. resistance decreases under applied field at all temperatures. We observe that with increasing field the value of negative MR increases, further with decreasing temperature the MR value increases. MR when sample is 300 K and 125K under an applied field of 50 kOe is $\sim$2 $\%$ and 22 $\%$ respectively. The large change in magnetoresistance in Pr$_2$CoMnO$_6$ material  is observed below PM-FM phase transition. The magnetic ordering below $T_c$ favors the transport of charge carriers through ordered spin channel, thus resistance fall under applied magnetic field. Further below $T_c$ with increasing field more and more spins align towards field direction and thus provides a favorable channel for charge conduction.  The variation of MR$\%$ vs $T$ under applied magnetic field of 50 kOe is shown in inset Fig. 8a. It is evident from figure that with decreasing temperature MR increases drastically. However, it is important there is sharp decrease in MR below FM ordering $T_c$, which conforms that the magnetic ordering plays a critical role in charge transport, which was earlier seen through a slope change in $\rho$(T,0) and $\rho$(T,H) data in Fig. 7b.

To further understand the charge conduction under applied field we attempt to analyze the field dependency of MR for Pr$_2$CoMnO$_6$. We found that above $T_c$ i.e. in paramagnetic state as well as at high field below $T_c$ MR scales well with $H^2$ which can be explained as Lorentz contribution. The magnetoresistance is best fitted to quadratic field dependency described as:\cite{ziman}

\begin{eqnarray}
\frac{\Delta\rho}{\rho} = \alpha (T) H^n
\end{eqnarray}
 where $\Delta$$\rho$ = $\rho$(B)-$\rho$(0), $H$ is the applied magnetic field and n takes the value 2

Fig. 8b shows the MR data plotted as $H^2$ vs MR $\%$ which shows the quadratic field dependency of magnetoresistance. Here it is quite evident from the figure that in PM phase the MR follows this model up to 30 kOe.  However, the MR collected in GP and in FM phase, at low field the MR seems to vary linear with applied magnetic field intensity $H$, whereas above 20 kOe the MR value again shows the quadratic field dependency. Solid lines in Fig. 8b are fitting due to Eq. 7 . It can be mentioned that negative MR with quadratic dependence on applied field has been observed for many other insulating materials such as, n-type CdSe, n-type GaAs etc.\cite{fran, blay, jin, imtiaz1, imtiaz2}] In these materials it is believed that the presence of quantum interference effect which drives a negative MR in such compounds. We have shed some light on the charge transport behavior in Pr$_2$CoMnO$_6$ double perovskite material, and we further believe that the subject is needed to be studied in more detail using much advanced local probing techniques.

\subsection{Conclusion}
In this work, we have reported the evolution of magnetism, Griffith phase and magnetotransport properties in 3$d$ based nano-crystalline double perovskite Pr$_2$CoMnO$_6$. We have successfully synthesized the nano-crystalline powder sample of Pr$_2$CoMnO$_6$ material. Structural analysis shows that the  sample is in orthorhombic crystal structure with pbnm phase group, with crystallite size $\sim$17 nm. The DC magnetization study reveals that the material shows a PM-FM phase transition around $T_c$ $\sim$173 K. The inverse magnetic susceptibility data shows a divergence from Curie-Weiss behavior around 206 K marked by $T_G$. Such down turn in $\chi^{-1}$ plot is signature of Griffiths phase evolution in PM state of a material. With detailed investigation of magnetic data we also found that Griffiths phase evolve in Pr$_2$CoMnO$_6$ well above FM ordering temperature $T_c$ in PM phase. we found the value of susceptibility exponent $\lambda$ well in agreement of GP model. Arrott plot analysis shows zero spontaneous moment between $T_c$ and $T_G$ which further conforms to the existence of Griffith phase. To understand the charge transport phenonmenon we have studied temperature dependent resistivity and magneto resistance in Pr$_2$CoMnO$_6$ sample. We found that the charge transport follows polarons model where hopping of small radius polarons involve in conduction process. The temperature dependency of resistivity shows a slope change around $T_c$ indicates spin polarons interaction in conduction process. Magnetoresistance study shows huge negative MR at low temperature and MR follows the quadratic field dependency which suggest that the quantum interference effect drive the negative MR in this material.\\

\section{Acknowledgment} 
  
We acknowledge AIRF (JNU) for magnetization and transport measurement, and thank Saroj Jha for the help in recording data. Author Ilyas Noor Bhatti acknowledge UGC for fellowship grant.


\begin{thebibliography}{}

\bibitem{ado} S. Rogado, J. Li, A. W. Sleight and M. A. Subramanian, Adv. Mater. (Weinheim, Ger.) \textbf{17}, 2225 (2005).
\bibitem{rov} I.N. Flerov, M. V. Gorev, K.S. Aleksandrov, A. Tressaud, J. Grannec and M. Couzi, Mater. Sci. Eng. \textbf{24}, 81 (1998).
\bibitem{son} M. Anderson, K. B. Greenwood, G. A. Taylor, and K. R. Poeppelmeier, Prog. Solid State Chem. \textbf{22}, 197 (993).
\bibitem{choudhury} D. Choudhury, P. Mandal, R. Mathieu, A. Hazarika, S. Rajan, A. Sundaresan, U. V. Waghmare, R. Knut, O. Karis, P. Nordblad, and D. D. Sarma, Phys. Rev. Lett., 108:127201, (2012).
\bibitem{kawa}  Y. Shimakawa, M. Azuma, and N. Ichikawa, Materials \textbf{4}, 153 (2011).
\bibitem{nair} H. S. Nair, D.A. Swain, N. Hariharan, S. Adiga, C. Narayana and S. Elzabeth, J. of Appl. Phys. \textbf{110}, 123919, (2011).
\bibitem{masud} G. Masud, A. Ghosh, J. Sannigrahi, and B. K. Chaudhuri, J. Phys.: Condens. Matter, \textbf{24},295902 (2012).
\bibitem{ali} A. O. Ayaş, Philosophical Magazine \textbf{98} No. 30, 2782 (2018).
\bibitem{ullah} M. Ullah, S. A. Khan, G. Murtaza, R. Khenata, N. Ullah and S. B. Omran, J. Magnet. and Magnet. Mater. \textbf{377} 197 (2015).
\bibitem{rto} M. Retuerto, A. Munoz, M. J. Martinez-Lope, J. A. Alonso,
F. J. Mompean, M.T. Fernandez-Diaz, and J. Sanchez-Benitez, Inorg. Chem.  \textbf{54}, 10890 (2015).
\bibitem{shi} R. Takahashi, I. Ohkubo, K. Yamauchi, M. Kitamura, Y. Sakurai, M. Oshima, T. Oguchi, Y. Cho and M. Lippmaa, Phys. Rev. B. \textbf{91}, 134107 (2015).
\bibitem{matte} D. Matte, M. de Lafontaine, A. Ouellet, M. Balli and P. Fournier, Phys. Rev. Applied \textbf{9}, 054042 (2018).
\bibitem{sco1} J. Blasco, J. García, G. Subías, J. Stankiewicz, J. A. Rodríguez-Velamazán, C. Ritter, J. L. Garcí -Muñoz, and F. Fauth, Phys. Rev. B \textbf{93}, 214401 (2016).
\bibitem{sco2} J. Blasco, G. Subías, J. García, J. Stankiewicz, J. A. Rodríguez-Velamazán, C. Ritter, and J. L. García-Muñoz, Solid State Phenomena \textbf{257}, 95 (2017).
\bibitem{hoo} R. C. Sahoo, Sananda Das, and T. K. Nath, J. Appl. Phys. \textbf{124}, 103901 (2018).
\bibitem{wang} X. L. Wang, J. Horvat, H. K. Liu, A. H. Li, S. X. Dou, Solid State Commun. \textbf{118}, 27 (2001).
\bibitem{khemchand} Khemchand Rawat, Meenakshi and Rabindra Nath Mahato, Mater. Res. Express \textbf{5},066110 (2018).
\bibitem{lshi} L. Shi, W. Liu, J. Zhao, Y. Li, S. Zhou, Y. Guo and Y. Wang, Mater. Res. Express \textbf{2} 076104 (2015).
\bibitem{liu} W. Liu, L. Shi, S. Zhou, J. Zhao, Y. Li and Y. Guo, J. Appl. Phys. \textbf{116}, 193901 (2014).
\bibitem{griffiths} R. B. Griffiths, Phys. Rev. Lett. \textbf{23}, 17 (1969).
\bibitem{zho} S. Zhou, Y. Guo, J. Zhao, L. He, and L. Shi, J. Phys. Chem. C \textbf{115}, 1535 (2011).
\bibitem{mandal} P. R. Mandal and T. K. Nath, Mater. Res. Express \textbf{2}, 066101 (2015).
\bibitem{jiang} W.J. Jiang, X. Z. Zhou, G. Williams, Phys. Rev. B \textbf{76}, 092404 (2007).
\bibitem{ang} W. J. Jiang, X. Z. Zhou,  G. Williams,  Y. Mukovskii,  K. Glazyrin, Phys. Rev. B  \textbf{77}, 064424 (2008).
\bibitem{chan} P. Y. Chan, N. Goldenfeld, M. Salamon, Phys. Rev. Lett. \textbf{97}, 137201 (2006).
\bibitem{zhou} S. M. Zhou, S. Y. Zhao, Y. Q. Guo, J. Y. Zhao, L. Shi, J. Appl. Phys. \textbf{107}, 033906 (2010).
\bibitem{mon} M. B. Salamon, S. H. Chun, Phys. Rev. B \textbf{68}, 014411 (2003).
\bibitem{akp1} A. K. Pramanik and A. Banerjee, Phys. Rev. B \textbf{81}, 024431 (2010).
\bibitem{jiang2} W. Jiang, X. Z. Zhou and G. Williams, Europhys. Lett. \textbf{84}, 47009 (2008).
\bibitem{akp2} A. K. Pramanik and A. Banerjee, J. Phys.: Condens. Matter \textbf{28}, 35LT02 (2016).
\bibitem{imtiaz1} Imtiaz Noor Bhatti, R. Rawat, A. Banerjee and A. K. Pramanik, J. Phys.: Condens. Matter \textbf{27}, 016005  (2014).
\bibitem{imtiaz2} Imtiaz Noor Bhatti R. S. Dhaka and A. K. Pramanik, Phys. Rev. B \textbf{96}, 144433 (2017).
\bibitem{magen} C. Magen, P. A. Algarabel, L. Morllon, J. P. Araujo, C. Ritter, M. R. Ibarra, A. M. Pereira and J. B. Sousa, Phys. Rev. Lett. \textbf{96}, 1167201 (2006).
\bibitem{neifel} E. A. Neifel’d, V. E. Arkhipov, N. A. Ugryumova, A. V. Korolev and Ya. M. Mukovskii, Bull. Russ. Acad. Sci. Phys. \textbf{71}, 1559 (2007). 
\bibitem{ziman} J. Ziman, \textit{Electrons and Phonons: The Theory of Transport Phenomena in Solids} Reprint: Clarendon Press, Oxford, (2007).
\bibitem{fran} O. Faran and Z. Ovadyahu, Phys. Rev. B \textbf{38}, 5457 (1988).
\bibitem{blay} F. Tremblay, M. Pepper, D. Ritchie, D. C. Peacock, J.E. F. Frost and G. A. C. Jones, Phys. Rev. B \textbf{39}, 8059 (1989).
\bibitem{jin} H. Jin, H. Jeong, T. Ozaki and J. Yu, Phys. Rev. B \textbf{80}, 075112 (2009).
\end{thebibliography}
\end{document}